\documentclass{llncs}
\usepackage{tabularx}
\usepackage{xcolor}
\usepackage{graphicx,float,epstopdf}
\usepackage{url}
\usepackage{bbold}
\usepackage{wrapfig}
\begin{document}\newcommand{\enzo}[1]{{\textcolor{red}{[#1]}}}

\title{Orthogonal Ensemble Networks\\ for Biomedical Image Segmentation}

 \author{Agostina J. Larrazabal$^1$, César Martínez$^1$, Jose Dolz$^2$, Enzo Ferrante$^1$}
 \institute{$^1$Research institute for signals, systems and computational intelligence, sinc(i), FICH-UNL / CONICET, Santa Fe, Argentina\\
 $^2$Laboratory for Imagery, Vision and Artificial Intelligence, École de Technologie Supérieure, Montreal, Canada}

\titlerunning{Orthogonal Ensemble Networks for Biomedical Image Segmentation}

\maketitle

\begin{abstract}

Despite the astonishing performance of deep-learning based approaches for visual tasks such as semantic segmentation, they are known to produce miscalibrated predictions, which could be harmful for critical decision-making processes. Ensemble learning has shown to not only boost the performance of individual models but also reduce their miscalibration by averaging independent predictions. In this scenario, model diversity has become a key factor, which facilitates individual models converging to different functional solutions. In this work, we introduce Orthogonal Ensemble Networks (OEN), a novel framework to explicitly enforce model diversity by means of orthogonal constraints. The proposed method is based on the hypothesis that inducing orthogonality among the constituents of the ensemble will increase the overall model diversity. We resort to a new pairwise orthogonality constraint which can be used to regularize a sequential ensemble training process, resulting on improved predictive performance and better calibrated model outputs. We benchmark the proposed framework in two challenging brain lesion segmentation tasks --brain tumor and white matter hyper-intensity segmentation in MR images. 
The experimental results show that our approach produces more robust and well-calibrated ensemble models 
and can deal with challenging tasks in the context of biomedical image segmentation.

\end{abstract}

\section{Introduction}
In the past few years, deep learning-based methods have become the \textit{de facto} solution for many computer vision and medical imaging tasks.
Nevertheless, despite their success and great ability to learn highly discriminative features, they are shown to be poorly calibrated \cite{guo2017calibration}, often resulting in over-confident predictions. 
This results in a major problem, which can have catastrophic consequences in critical decision-making systems, such as medical diagnosis, where the downstream decision depends on predicted probabilities.

Ensemble learning is a simple strategy to improve both the robustness and calibration performance of predictive models \cite{lakshminarayanan2016simple,stickland2020diverse}. In this scenario, a common approach is to train the same model under different conditions, which can foster the model convergence to different functional solutions. Techniques to produce ensembles include dataset shift \cite{ovadia2019can}, Monte-Carlo Dropout \cite{gal2016dropout}, batch-ensemble \cite{wen2020batchensemble} or different model hyperparameters \cite{wenzel2020hyperparameter}, among others. Then, by averaging the predictions, individual mistakes can be dismissed leading to a reduced miscalibration. In this context, ensuring \textit{diversity} across models is a key factor to build a robust ensemble. To promote model diversity in ensembles many mechanisms have been proposed. These include using latent variables \cite{sinha2020dibs}, integrating attention in the embeddings to enforce different learners to attend to different parts of the object \cite{kim2018attention} or isolating the adversarial vulnerability in
sub-models by distilling non-robust features to induce diverse outputs against a transfer attack \cite{yang2020dverge}.

Nevertheless, despite the relevance of obtaining well-calibrated models in clinical applications, relatively few works have studied this problem. 
Particularly, in the context of medical image segmentation, it was suggested that models trained with the well-known soft Dice loss \cite{milletari2016v} produce miscalibrated models \cite{sander2019towards}, which tend to be highly overconfident.
Furthermore, the recent work in \cite{mehrtash2020confidence} proposed the use of ensembles to improve confidence calibration. 
However, the importance of model diversity was not assessed in this work. Thus, given the negative impact of miscalibrated models in health-related tasks, and the current practices in medical image segmentation of systematically employing the Dice loss as an objective function, we believe it is of paramount importance to investigate the effect of ensemble learning in image segmentation, and how to enforce model diversity to generate high-performing and well-calibrated models.\\

\noindent \textbf{Contributions.} In this work, we propose a novel learning strategy to boost model diversity in deep convolutional neural networks (DCNN) ensembles, which improves both segmentation accuracy and model calibration in two challenging brain lesion segmentation scenarios. 
The main hypothesis is that inducing orthogonality among the constituents of the ensemble will increase the overall model diversity. We resort to a novel pairwise orthogonality constraint which can be used to regularize a sequential ensemble training process, resulting on improved predictive performance and better calibrated model outputs. In this context, our contributions are 3-fold: (1) we propose a novel filter orthogonality constraint for ensemble diversification, (2) we show that diversified ensembles improve not only segmentation accuracy but also confidence calibration and (3) we showcase the proposed framework in two challenging brain lesion segmentation tasks, including tumor and white-matter hyperintensity (WMH) segmentation on magnetic resonance images.

\vspace{-1mm}

\section{Related works} 

Diversifying ensembles has been used 
to improve classification and segmentation performance of DCNNs in several contexts. In \cite{kamnitsas2017ensembles} authors propose an explicit way to construct diverse ensembles bringing together multiple CNN models and architectures. Although they obtain successful results, this approach requires to manually design and train various architectures. An ensemble of 3D U-Nets with different hyper-parameters for brain tumor segmentation is proposed in \cite{feng2020brain}, where authors point out that using different hyper-parameters reduces the correlations of random errors with respect to homogeneous configurations. However, no study on the diversity of the models and its influence on performance is presented. 
In \cite{ma2021ensembling} authors present a different view, highlighting that many automatic segmentation algorithms tend to exhibit asymmetric errors, typically producing more false positives than false negatives. By modifying the loss function, they train a diverse ensemble of models with very high recall, while sacrificing their precision, with a sufficiently high threshold to remove all false positives. While the authors achieve a significant increase in performance 
no study on the final calibration of the ensemble is carried out.
 
Following the success of ensemble methods at improving discriminative performance, its capability to improve confidence calibration has begun to be explored. \cite{lakshminarayanan2016simple} uses a combination of independent models to reduce confidence uncertainty by averaging predictions over multiple models. In \cite{mehrtash2020confidence} authors achieve an improvement in both segmentation quality and uncertainty estimation by training ensembles of CNNs with random initialization of parameters and random shuffling of training data. While these results are promising, we believe that confidence calibration can be further improved by directly enforcing diversity into the models instead of randomly initializing the weights. 
 
As pointed out in \cite{wang2020orthogonal} over-sized DNNs often result in a high level of overfitting and many redundant features. However, when filters are learned to be as orthogonal as possible, they become decorrelated and their filter responses are no longer redundant, thereby fully utilizing the model capacity. \cite{ayinde2019regularizing} follows a very similar approach but they regularize both negatively and positively correlated features according to their differentiation and based on their relative cosine distances. Differently from these works where ortogonality constraints are used to decorrelated the filters within a single model, here we propose to enforce filter orthogonality among the constituents of the ensemble to boost model diversity.\\


\vspace{-4mm}

\section{Orthogonal Ensemble Networks for Image Segmentation}
Given a dataset $\mathcal{D} = \{(\mathbf{x}, \mathbf{y})_i\}_{0 \leq i \leq |\mathcal{D}|}$ composed of images $\mathbf{x}$ and corresponding segmentation masks $\mathbf{y}$, we aim at training a model which approximates the underlying conditional distribution $p(\mathbf{y}|\mathbf{x})$, mapping input images $\mathbf{x}$ into segmentation maps $\mathbf{y}$. Thus, $p(y_j = k|\mathbf{x})$ will indicate the probability that a given pixel (or voxel) $j$ is assigned class $k \in \mathcal{C}$ from a set of possible classes $\mathcal{C}$. The distribution is commonly approximated by a neural network $f_\mathbf{w}$ parameterized by weights $\mathbf{w}$. In other words, $f_\mathbf{w}(\mathbf{x}) = p(\mathbf{y}|\mathbf{x}; \mathbf{w})$. Parameters $\mathbf{w}$ are learnt so that
they minimize a particular loss function over the training dataset. Given a set of segmentation networks $\{ f_{\mathbf{w}^1}, f_{\mathbf{w}^2} ... f_{\mathbf{w}^N}\}$, a simple strategy to build an ensemble network $f_\mathbf{E}$ is to average their predictions as:

\begin{equation}
    f_\mathbf{E}(\mathbf{x}) = \frac{1}{N} \sum_{i=1}^N f_\mathbf{w^i}(\mathbf{x}).
\end{equation}

Under the hypothesis that diversifying the set of models $f_\mathbf{w}^i$ will lead to more accurate and calibrated ensemble predictions, we propose to boost its overall performance by incorporating pairwise orthogonality constraints during training.\\

\textbf{Inducing model diversity via orthogonal constraints.}
Modern deep neural networks are parameterized by millons of learnable weights, resulting in redundant features that can be either a shifted version of each other or be very similar with almost no variation \cite{ayinde2019regularizing}. Inducing orthogonality between convolutional filters from the same layer of a given network has shown to be a good way to reduce filter redundancy \cite{wang2020orthogonal}. Here we exploit this principle not only to avoid redundancy within a single neural model, but among the constituents of a neural ensemble. 

Given two vectors $\textbf{x}$ and $\textbf{y}$, cosine similarity quantifies orthogonality (or decorrelation), ranging from -1 (i.e., exactly opposite) to 1 (i.e., exactly the same), with 0 indicating orthogonality. It can be defined as:

\begin{equation}
    \mathrm{SIM}_C(\textbf{x},\textbf{y})=\frac{<\textbf{x},\textbf{y}>}{||\textbf{x}||\ ||\textbf{y}||}.
\end{equation}

Following \cite{ayinde2019regularizing}, we consider the squared cosine similarity to induce orthogonality between filters through a new regularization term in the loss function. An advantage of this measure is that it takes into account both negative and positive correlations.

In order to enforce diversity within and between the ensemble models, we propose to include two regularization terms into the overall learning objective. The first one, referred to as self-orthogonality loss ($\mathcal{L}_{\mathrm{SelfOrth}}$), aims at penalizing the correlation between filters in the same layer, for a given model. Thus, for a given convolutional layer $l$, this term is calculated as follows:

\begin{equation}\label{eq:selforth}
   \mathcal{L}_{\mathrm{SelfOrth}}(\textbf{w}_l) =  \frac{1}{2} \sum_{i=1}^{n} \sum_{j=1,j \neq i}^{n}\mathrm{SIM}_C(\textbf{w}_{l,i},\textbf{w}_{l,j})^2,
\end{equation}
where $\textbf{w}_{l,i}$ and $\textbf{w}_{l,j}$ are vectorized versions of each of the $n$ convolutional kernels from layer $l$. 
We also define an inter-orthogonality loss term ($\mathcal{L}_{\mathrm{InterOrth}}$) which penalizes correlation between filters from different models in the ensemble. To this end, following a sequential training scheme, the inter-orthogonality loss for layer $l$ of model $N_e$ is estimated as follows:

\begin{equation}
   \mathcal{L}_{\mathrm{InterOrth}}(\textbf{w}_l;\{\textbf{w}^e_l\}_{0\leq e<N_e}) = \frac{1}{N_e} \sum_{e=0}^{N_e-1}  \sum_{i=1}^{n} \sum_{j=1}^{n} \mathrm{SIM}_C(\textbf{w}_{l,i},\textbf{w}^e_{l,j})^2,
\end{equation}
where $\{\textbf{w}^e_l\}_{0\leq e<N_e}$ are the parameters of the previous $N_e-1$ models trained during the sequential ensemble construction.\\

Thus, the learning objective to train the proposed OEN amounts to:

\begin{equation}\label{eq:_selfloss}
   \mathcal{L} = \mathcal{L}_{Seg} + \lambda \sum_{l} \Big(\mathcal{L}_{\mathrm{SelfOrth}}(\textbf{w}_l) +  \mathcal{L}_{\mathrm{InterOrth}}(\textbf{w}_l;\{\textbf{w}^e_l\})\Big),
\end{equation}
where $\mathcal{L}_{Seg}$ is the segmentation loss (e.g. soft Dice loss or cross entropy) and $\lambda$ is a hyperparamether controlling the influence of the orthogonality terms. \footnote{Our code associated to the orthogonal ensemble networks training is publicly available at: \url{https://github.com/agosl/Orthogonal_Ensemble_Networks}}\\ 

\vspace{-4mm}
 \section{Experimental framework}
 
\textbf{Database description.} We benchmark the proposed method in the context of brain tumor and WMH segmentation in MR images. For brain tumor we use the BraTS 2020 dataset \cite{Brats_bakas2017advancing,Brats_bakas2018identifying,Brats_menze2014multimodal} which contains 369 images with expert segmentation masks (including GD-enhancing tumor, peritumoral edema, and the necrotic
and non-enhancing tumor core). Each patient was scanned with FLAIR, T1ce, T1, and T2. 
The images were re-sampled to an isotropic 1.0$mm$ voxel spacing, skull-striped and co-registered by the challenge organizers. 
The provided training set, we divide the database in training (315), validation (17) and test (37). The second dataset \cite{kuijf2019standardized} consists of 60 MR images with binary masks indicating the presence of WMH lesions. For each subject, co-registered 3D T1-weighted and a 2D multi-slice FLAIR images were provided. We split the dataset in training (42), validation (3) and test (15). All images have 3$mm$ spacing in the z dimension, and approximately 1$mm$ $\times$ 1$mm$ in the axial plane.\\

\noindent \textbf{Segmentation network.} For all the experiments, the backbone segmentation network was a state-of-the-art ResUNet architecture \cite{zhang2018road} 
implemented in Keras 2.3 with TensorFlow as backend, with soft Dice \cite{milletari2016v} as segmentation loss $\mathcal{L}_{Seg}$. 
For the BraTS dataset, the input was a four-channel tensor (FLAIR, T1ce, T1, and T2) and a softmax activation was used as output, whereas a two-channel input (T1, FLAIR) was employed in the WMH, with a sigmoid activation function in the output. During training, patches of size $64\times64\times64$ were extracted from each volume, and networks were trained until convergence by sampling the patches randomly, with equal probability for each class in the case of tumour segmentation, and 0.9 probability in the case of WMH. We used Adam optimizer with a batch size of 64. The initial learning rate was set to 0.001 for BraTS and 0.0001 for WMH, and it was reduced by a factor of 0.85 every 10 epochs. Hyper-parameters were chosen using the validation split, and results reported on the hold-out test set. \\

\noindent \textbf{Baselines and ensemble training.} We trained two different baselines to benchmark the proposed method. In the first one (\emph{random} ensemble) each model was randomly initialized and trained to reduce only the segmentation error $\mathcal{L}_{Seg}$. Therefore, its main source of diversity comes from the initialization of the weights. 
The second approach (\emph{self-orthogonal} ensemble) includes the $\mathcal{L}_{\mathrm{SelfOrth}}$ term in the learning objective, creating and ensemble of models individually trained with the self-orthogonality constraint. Thus, while each model learns orthogonal filters, orthogonality between different models in the ensemble was not imposed. We compared these two models with the proposed orthogonal ensemble network which also encourages inter-model diversity by minimizing the full objective defined in Eq. \ref{eq:_selfloss} (referred as \textit{inter-orthogonal}). Note that in our approach models are trained sequentially. For each of the proposed settings we trained 10 models. During evaluation, we assembled groups of 1, 3 and 5 models from each setting by averaging the individual probability outputs. To provide better statistics, we repeated this process 10 times, each with different model selection. We empirically observed that beyond 5 models, the performance of the ensemble did not improve. Furthermore, $\lambda$ was set to 0.1 and 1 for the WMH and brain tumour segmentation task, respectively.\\


\noindent \textbf{Measuring calibration for image segmentation.} 
Given a segmentation network $f_\mathbf{w}$, if the model is well-calibrated its output for a single pixel $j$ can be interpreted as the probability $p(y_j = k|\mathbf{x}; \mathbf{w})$ for a given class $k \in \mathcal{C}$. In this case, the class probability can be seen as the model confidence or probability of correctness, and can be used as a measure for predictive uncertainty at the pixel level \cite{mehrtash2020confidence}. 
A common metric used to measure calibration performance is the Brier score \cite{brier1950verification}, a proper scoring rule whose optimal value corresponds to a perfect prediction. In other words, a system that is both perfectly calibrated and perfectly discriminative will have a Brier score of zero. In the context of image segmentation, for an image with $N$ pixels (voxels), the Brier score can be defined as:

\begin{equation}
    Br = \frac{1}{N} \sum_{i=1}^{N}  \frac{1}{|\mathcal{C}|} \sum_{k=1}^{|\mathcal{C}|} \Big( p(y_i = k |\mathbf{x}; \mathbf{w}) - \mathbb{1} [\bar{y}_{i} = k]  \Big)^2,
\end{equation}
where $\mathbb{1} [\bar{y}_{i} = k]$ is the indicator function whose value is 1 when $\bar{y}_i$ (the ground truth class for pixel $i$) is equal to $k$, and 0 otherwise. \\

\noindent \textbf{Stratified Brier Score.} In problems with highly imbalanced classes (such as brain lesion segmentation where most of the pixels are background), calibration may be good overall but poor for the minority class. In this case, the majority class will dominate and miscalibration in the class of interest will not be reflected in the standard Brier score. In \cite{wallace2014improving}, the authors proposed the 
stratified Brier score to measure calibration in binary classification problems with high imbalance. Here, we extend this concept to the segmentation task and propose to measure the stratified Brier score individually per-class, treating every structure of interest as a binary segmentation problem, to account for mis-calibration in the minority classes. For a given image with ground truth segmentation $\mathbf{\bar{y}}$, we construct the \textit{stratified} Brier score for the class $k$, $Br^k$, by computing it only in the subset of pixels $\mathcal{P}_k = \{p: \bar{y}_p = k\}$, i.e. pixels whose ground truth label is $k$. The problem is therefore binarized considering all the other classes within a single background class. The formulation of the stratified Brier score $Br^k$ is given by:

\begin{equation}
    Br^k = \frac{1}{|\mathcal{P}_k|} \sum_{i \in \mathcal{P}_k} \Big( p(y_i = k |\mathbf{x}; \mathbf{w}) - \mathbb{1} [\bar{y}_{i} = k]  \Big)^2.
\end{equation}

\noindent \textbf{Segmentation evaluation.} In addition to the metrics presented to measure the model miscalibration, we resort to the common Dice Similarity coefficient (DSC) 
to assess the quality of the segmentations.

\begin{figure}[h!]
   \includegraphics[width=\textwidth]{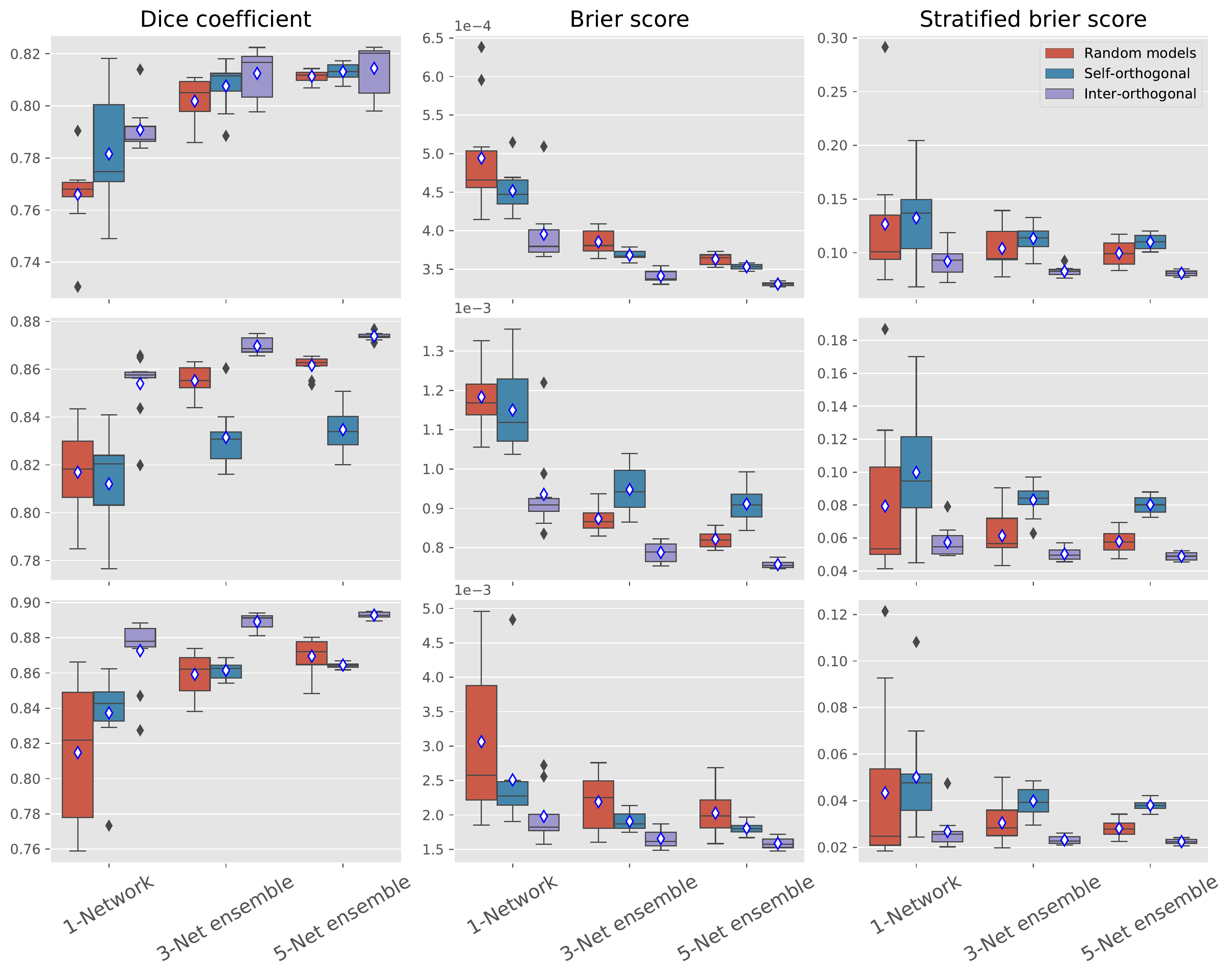}
        \caption{Quantitative evaluation of the proposed method on BraTS: Rows from top to bottom show results for: (i) enhanced tumor; (ii) tumor core; (iii) whole tumor. 
        Boxplots show mean and standard deviation for predictions obtained with individual models, 3-networks ensembles and 5-networks ensembles.}
    \label{fig:boxplots_brats}   
\end{figure}

\vspace{-9mm}
\section{Results and discussion}

\begin{figure}[t!]
   \includegraphics[width=\textwidth]{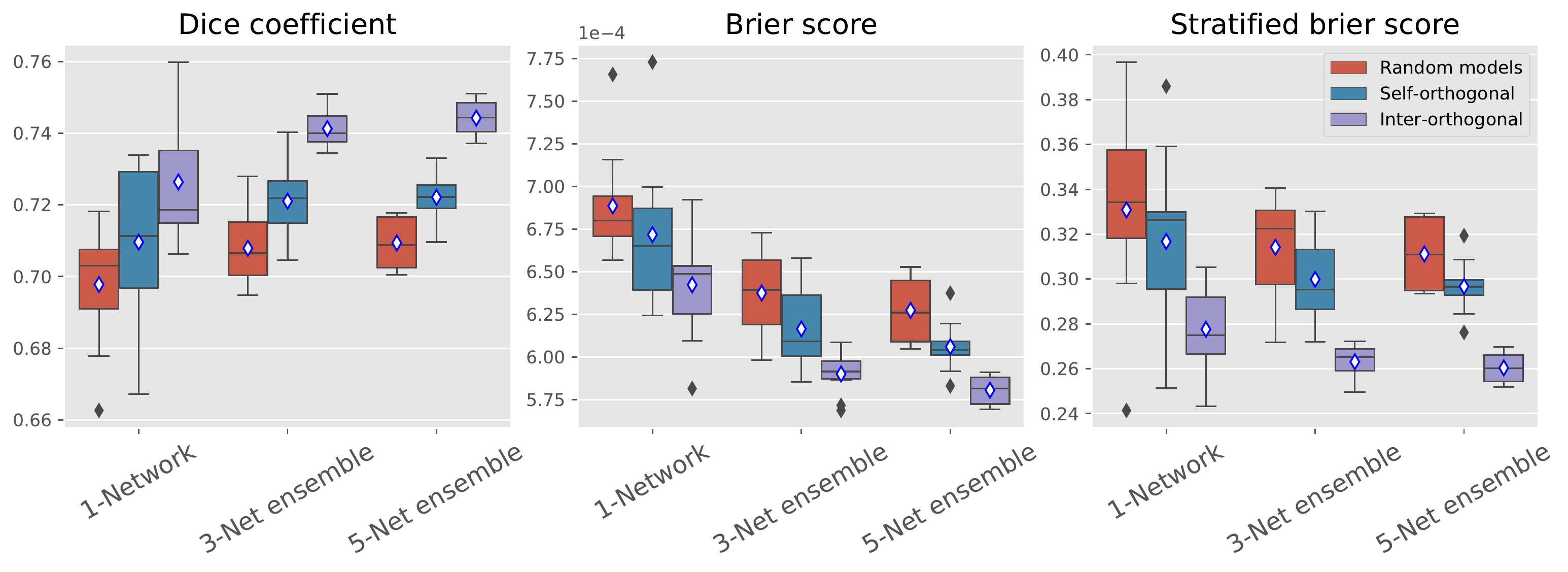}
        \caption{Quantitative evaluation of the proposed method for WMH segmentation. 
        Boxplots show mean and standard deviation for predictions obtained with individual models, 3-networks ensembles and 5-networks ensembles.}
    \label{fig:boxplots_wmh}   
\end{figure}

We present quantitative results for brain tumor and WMH segmentation in Fig. \ref{fig:boxplots_brats} and Fig. \ref{fig:boxplots_wmh}, respectively. We can observe that the model just integrating \textit{self-orthogonality} outperforms the baseline model across groups and metrics. This improvement is further stressed when explicitly enforcing model diversity by incorporating the $inter-orthogonality$ term computed between pairs of models during sequential training. In particular, our proposed learning strategy consistently leads to improvement on both model calibration and segmentation performance and across the two different segmentation tasks. This demonstrates the benefits of the proposed learning strategy to generate well-calibrated and highly performing segmentation models.

Another important observation is related to differences between Brier and stratified Brier scores. Given the small Brier value reported for all the models (less than $1^{-3}$), one could think that these models are well calibrated. However, when having a closer look at the stratified Brier score, the higher value (more than 0.1 in most of the cases) reflects calibration issues. This results from the majority class dominating the traditional Brier score. Thus, studying the stratified Brier score allows us to better appreciate the improvements obtained by the inter-orthogonal ensemble with respect to the other models. 


\begin{figure}
\begin{center}

   \includegraphics[scale=0.5]{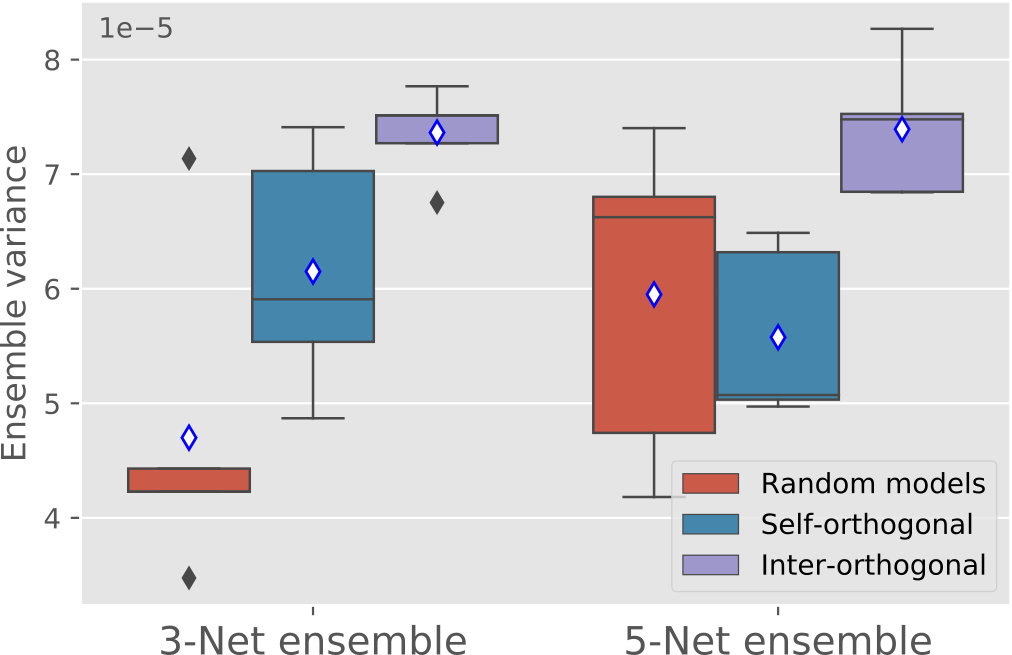}
    \caption{Quantitative evaluation of the ensembles diversity. Boxplots depict the mean and standard deviation of the variance in the predictions when training the ensemble with and without the proposed orthogonal losses.}
    \label{fig:boxplots_variance}
    \end{center}
\end{figure}

 In addition, we depict in Fig. \ref{fig:boxplots_variance} the variance in the predictions across the components of the ensemble trained with and without the orthogonal losses, demonstrating that the orthogonal constraints bring diversity to the ensemble. As expected, we found that integrating the inter-orthogonal objective term leads to an increase in the variance of the predictions compared to the baseline models.

Last but not least, it is surprising to see that the inter-orthogonal regularization term boosts the performance even when considering the individual models. We believe that this is due to a regularization effect of the inter-orthogonal term, which implicitly reduces the complexity of the model by adding orthogonality constraints with respect to specific points in the parameter space, i.e. the weights of the previously trained models.
\vspace{-4mm}

\section{Conclusions}

In this work we introduced Orthogonal Ensemble Networks (OEN), a novel training framework that produces more diverse ensembles. Our formulation explicitly imposes orthogonal constraints during training by integrating a regularization term that enhances the inter-model diversity. Experiments across two different segmentation tasks have demonstrated that, in addition to improved segmentation performance, the proposed inter-model orthogonality constraints reduce miscalibration, leading to more reliable predictions.

\section*{Acknowledgments}
 The authors gratefully acknowledge NVIDIA Corporation with the donation of the GPUs used for this research, and the support of UNL (CAID-0620190100145LI, CAID-50220140100084LI) and ANPCyT (PICT 2018-03907). This research was enabled in part by support provided by Calcul Québec and Compute Canada.

%
%
\bibliographystyle{splncs}
\bibliography{bibliography.bib}

\end{document}